\documentclass[referee]{aa}
\usepackage[dvips]{graphicx}
\usepackage{amsfonts}
\begin{document}
\title{Rationale for the use of color information on Eddington}
\author{P.~Bord\'e\inst{1} \and A.~L\'eger\inst{2} \and D.~Rouan\inst{1}
        \and A.~C.~Cameron\inst{3}}
\offprints{P.~Bord\'e}
\institute{LESIA, UMR8109, Observatoire de Paris, 5 place Jules Janssen,
           F-92195 Meudon CEDEX, France \\
           \email{Pascal.Borde@obspm.fr, Daniel.Rouan@obspm.fr}
           \and
           IAS, UMR8617, Universit\'e Paris-Sud, F-91405 Orsay, France \\
           \email{Alain.Leger@ias.fr}
           \and
           University of St Andrews, North Haugh, St Andrews, Scotland KY16 9SS \\
           \email{Andrew.Cameron@st-and.ac.uk}
           }
\date{Received ; accepted }
\abstract{
For the Eddington mission, the intrinsic stellar variability can be a major
source of noise in the detection of extrasolar planets by the transit method.
We derive that most detections of terrestrial planets (1--2~R$_\oplus$) will
occur around G or K stars with 15--16th magnitude. When these stars are 7--12
times more variable than the Sun on a 10 hour timescale, we demonstrate that the
detection can be performed with a higher S/N provided composite lightcurves
obtained with the combination of two colors are used instead of white ones. The
level of 10 hour variability for K stars is quite uncertain. We make two
``guess-estimates'' of it and find that it could be several times larger than
the solar value. If these estimates were relevant, the color information would
not provide a significant advantage. Although we do not demonstrate a need for
colors, \emph{we point out the risk of an unpleasant surprise regarding the 10
hour stellar variability. Indeed, there is presently no qualified proxy for this
variability.} Besides, if Eddington were designed to provide this information at
the cost of added complexity but not sensitivity, white photometry by channel
summation would still be as efficient. Considering the risk that 10 hour
variability is higher than estimated, \emph{the Precaution Imperative points to
a study of practical implementations of photometry in different colors} before
taking irreversible decisions about the Eddington instrument.   
}
\maketitle
%
%
\section{Introduction}
The purpose of this research note is to calculate the gain in detectability of
extrasolar planets brought by the use of color information instead of the white
flux alone, in the case of the Eddington mission (Favata et al. \cite{favata}).

The monitoring of the solar variability performed by SOHO/SPM in three bands
(Fr\"ohlich et al. \cite{frohlich}) shows that there is a strong correlation
between the different colors, the variation in the blue being significantly
larger than in the red. On the contrary, a transit is essentially perceived as
an achromatic variation of the stellar flux. Thus, a properly weighted linear
combination of the colored lightcurves can provide a composite lightcurve almost
free of stellar variability. Unfortunately, this has to be traded off against 
an increase of quantum noise. The following calculation evaluates the
variability threshold above which the signal to noise ratio (S/N) is increased
by the variability subtraction.
%
%
\section{White detection S/N}
Let us consider the problem from the standpoint of pure detection. Then, the
highest S/N is achieved when the lightcurve is averaged over the transit duration,
that is to say transits are reduced to one point. The noise figure for Eddington
is expected to be dominated by the quantum noise with an additional contribution
of the stellar variability noise. If $N_e$ is the total number of photoelectrons
collected during a transit, $\sigma_\star\,N_e$ the standard deviation of
the stellar variability, and considering that these two noises as uncorrelated,
the white light S/N on a single transit is
	\begin{equation}
	(S/N)_\mathrm{w} = {\left(\frac{R_\mathrm{p}}{R_\star}\right)}^2
	      \sqrt{\frac{N_e}{1+\sigma_\star^2\,N_e}},
	\end{equation}
where $R_\mathrm{p}$ and $R_\star$ are the radii of the planet and of its star,
respectively.
%
%
\section{Colored detection S/N}
\begin{figure}
	\centering
	\includegraphics[width=8cm]{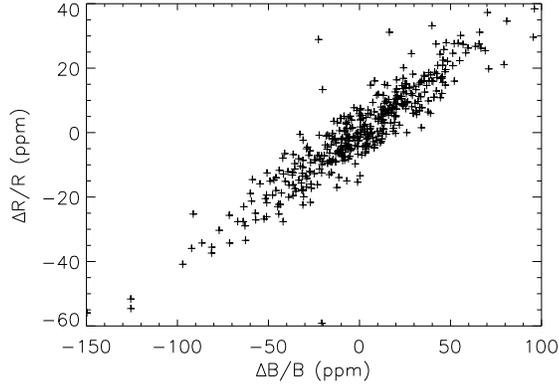}
	\caption{Correlation between the solar relative variabilities in the red
	(862~nm) and blue (402~nm) narrow filters of SOHO/SPM, at the timescale of a
	few hours. The coefficient of correlation is 0.913.}
	\label{fig:spm}
	\end{figure}

Let us assume now that the stellar flux is observed in two colored channels,
denoted B (blue) and R (red), collecting the fractions $x_B$ and $x_R$ of the
total number of photoelectrons. As the stellar variations are highly correlated
at the timescale of a few hours (Fig.~\ref{fig:spm}), one can remove most of the
stellar variability noise by combining the relative colored flux variations:
	\begin{equation}
	S = \frac{\Delta B}{B} - k\,\frac{\Delta R}{R},
	\end{equation}
where $k$ is a constant adjusted to minimize $S$ when $\Delta B$ and $\Delta R$
are due to the stellar variability alone. Now, we assume that $S$ is due to the
transit of a planet in front of a star producing a total of $N_e$ photoelectrons.
Denoting by $N$ the new quantum noise, the S/N in colored light is
	\begin{equation}
	(S/N)_\mathrm{c} = (k-1) {\left(\frac{R_\mathrm{p}}{R_\star}\right)}^2
	      \sqrt{\frac{N_e}{x_B^{-1}+k^2\,x_R^{-1}}}.
	\end{equation}
Using the SOHO/SPM data and the transmission curve of COROT (Eddington's
precursor, Rouan et al. \cite{rouan}), one computes ${k=1.62}$, ${x_B = 32\%}$
and ${x_R = 42\%}$, as being the optimum values for the Sun.
%
%

%
\section{Magnitude and spectral type histograms of stars with expected detection}
\begin{figure*}
	\centering
	\includegraphics[width=12cm]{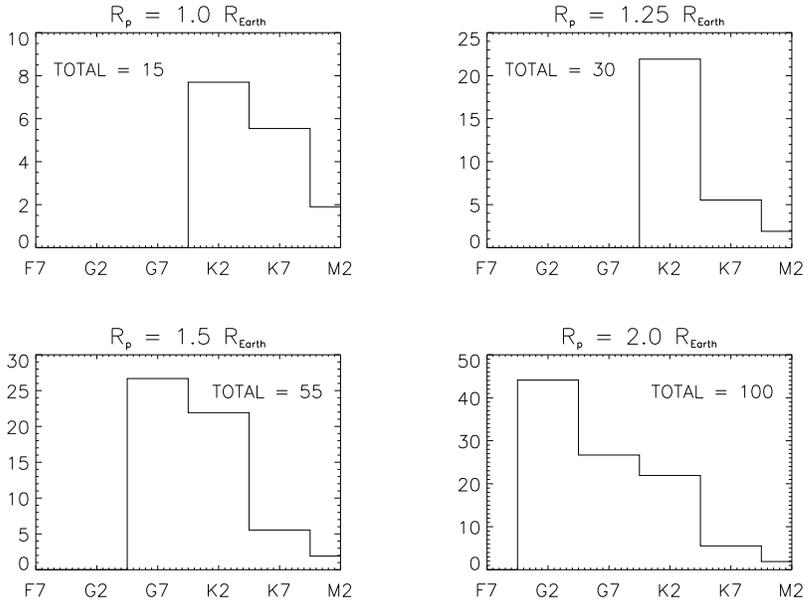}
	\caption{Histograms of expected detections vs. the parent star spectral
	type for $R_\mathrm{p}=1.0$, 1.25, 1.5 and 2.0~R$_\oplus$.
	Every star is assumed to be orbited by a planet of the given size at 278~K
	blackbody temperature (1~AU for a G2V star).}
	\label{fig:det_sp}
	\end{figure*}

\begin{figure*}
	\centering
	\includegraphics[width=12cm]{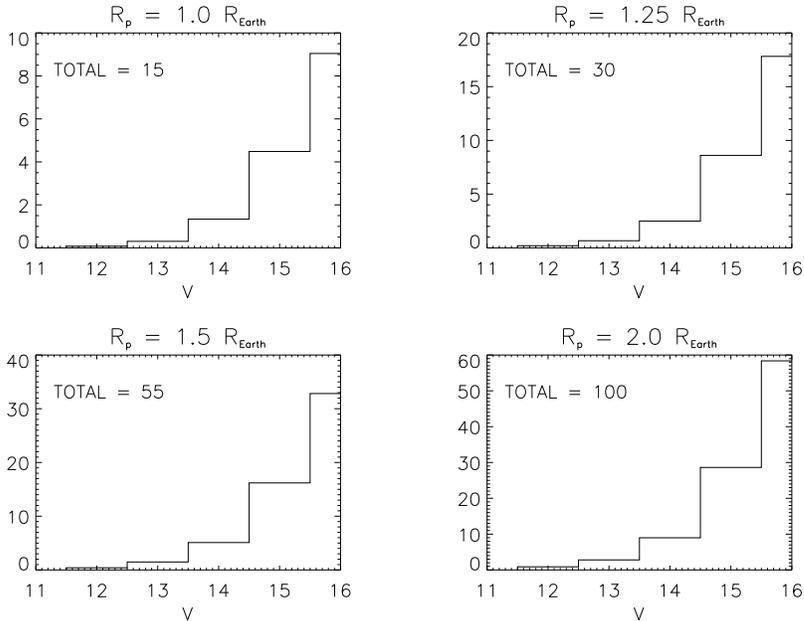}
	\caption{Same as Fig.~\ref{fig:det_sp} vs. the visual magnitude of the
	parent star.}
	\label{fig:det_mv}
	\end{figure*}

By using a simple transit detection algorithm (Bord\'e et al. \cite{borde01},
\cite{borde02}), we estimate, in the stellar field envisioned for
Eddington ($l_\mathrm{II}=70\degr$, $b_\mathrm{II}=5\degr$, 7.1~deg$^2$), the
number of stars with expected planet detections, as a function of their
spectral type (Fig.~\ref{fig:det_sp}) and magnitude (Fig.~\ref{fig:det_mv}).
These computations assume that every star is orbited by a planet with the given
radius at 278~K blackbody temperature (1~AU for a G2V star).
We conclude that for Earth-size planets most detections occur around K stars
with 15--16th magnitude. For larger planets, the detection peak shifts toward
G stars still with 15--16th magnitude.
%
%
\section{Variability threshold}
The variability threshold $\sigma_\star^\mathrm{min}$ above which one benefits
from the colored detection is given by the condition
	\begin{eqnarray}
	(S/N)_\mathrm{c} & \ge & (S/N)_\mathrm{w} \\
	\Longrightarrow \quad \sigma_\star^2 & \ge & \frac{1}{N_e}
	\left\{ \frac{x_B^{-1}+k^2\,x_R^{-1}}{(k-1)^2}-1 \right\} \\
	\mbox{i.e.} \quad \sigma_\star^\mathrm{min} & \approx &
	                  \frac{4.8}{\sqrt{N_e}}.
	\end{eqnarray}
Thus, the colored detection becomes more efficient than the white one when the
stellar variability noise overcomes the quantum noise by a factor of $\approx 5$.

Now, let us adopt the Astrium concept for Eddington: 4 Schmidt telescopes
with ${D=0.6}$~m pupils pointing at the same direction. We assume that all
telescopes are equiped with dichroic plates allowing the simultaneous measure
of blue and red signals. Instead of dichroic plates, one may use a dispersive
device, for instance derived from that by Chao \& Chi (\cite{chao}).
For a G2 star and a planet at 1~AU, Table~\ref{tab:min} lists the values
obtained for  $\sigma_\star^\mathrm{min}$ as a function of the star visual
magnitude, $m_V$. These values are converted into solar variability levels using
high-pass filtered SOHO/SPM data. At the time scale of a transit, 11.2~h for an
Earth-like planet, $\sigma_\odot \approx 6.3\,10^{-5}$.
\begin{table}[htbp]
\centering
\begin{tabular}{cccc}
\hline
$m_V$ & $N_e$ & $\sigma_\star^\mathrm{min}$ & $\sigma_\star^\mathrm{min}/\sigma_\odot$ \\
\hline
10 & $1.0\,10^{10}$ & $4.7\,10^{-5}$ & 0.8 \\
11 & $4.1\,10^9$    & $7.5\,10^{-5}$ & 1.2 \\
12 & $1.6\,10^9$    & $1.2\,10^{-4}$ & 1.9 \\
13 & $6.5\,10^8$    & $1.9\,10^{-4}$ & 3.0 \\
14 & $2.6\,10^8$    & $3.0\,10^{-4}$ & 4.7 \\
15 & $1.0\,10^8$    & $4.7\,10^{-4}$ & 7.5 \\
16 & $4.1\,10^7$    & $7.5\,10^{-4}$ & 11.9 \\
\hline
\end{tabular}
\caption{Stellar variability thresholds for a planet orbiting a G2 star at 1~AU.}
\label{tab:min}
\end{table}
%
%
%
\section{Estimate of stellar variability on a 10 hour timescale}
To address the question of stellar variability on a 10 hour timescale, the
\emph{only star} for which we have data with sufficient precision is the Sun, as
observed by SOHO. Work by the St Andrews group and co-workers (see for instance
http:\textbackslash \textbackslash capella.st-and.ac.uk\textbackslash
$\sim$acc4\textbackslash corot\_ao\_bid.pdf)
as well as Carpano, Aigrain \& Favata (2002) indicates that: 
\begin{description}
\item (i) the amplitude of variability on timescales less than one day does not 
vary strongly during the solar cycle; 

\item (ii) the color signature of the variability on these short timescales 
(from high-pass filtered VIRGO/SPM data) is indistinguishable from that of 
either spots or faculae.
\end{description}
This leads to the conclusion that whatever phenomenon produces the 
10 hour timescale variability, it does \emph{not} scale with the overall
coverage of active regions (solar spots) and therefore variability on a 10 day
timescale. 

SOHO/MDI continuum images taken around solar minimum have been examined,
searching for low-contrast features that might be responsible for this
variability. The only features that showed up were the bright network elements
dotted around the edges of supergranules (see http:\textbackslash
\textbackslash capella.st-and.ac.uk\textbackslash $\sim$jrb3\textbackslash
sunmovie3.gif).
They are limb-brightened in the same way as faculae, showing up strongly only
when they are near the limb. If they are the source of the hour timescale
variability, this explains the color signature very nicely. Since the network
elements appear to be a ``magnetic carpet'' phenomenon, this would also explain
why the amount of variability on these timescales does not vary much during the
solar cycle, a distinct physical process. 

Now, how can these preliminary conclusions be extrapolated to other stars? One
would predict that if the hour timescale variability is a granulation/network
phenomenon, its amplitude would not change much from year to year, even for
more active stars. 

\emph{The estimate of the variability amplitude is much more uncertain},
especially for latter type stars where the contrast of the filament network to
quiet photosphere can be higher. In fact, \emph{there is no qualified proxy for
the 10 hour stellar activity}. For any star, the luminosity 10 hour
variability depends upon the total number of filaments, $n$, and their
individual brightness, $\delta$, as a function $f(n,\delta)$. For instance, if
most of the variation comes from Poissonian fluctuations of $n$,
$f(n,\delta) = \sqrt{n} \, \delta$.

Let us consider two estimates: 
\begin{enumerate}
\item One can assume that the number and
brightness of network elements is independent of spectral type, i.e.
$n = n_\odot$, $\delta = \delta_\odot$ and $f = f_\odot$. Then the standard
deviation of the relative change of the stellar luminosity is
	\begin{equation}
	\sigma_{\star,1} = \frac{f(n_\odot,\delta_\odot)}{L_\star}
                 = \sigma_\odot \, \frac{L_\odot}{L_\star}.
	\end{equation}
The corresponding values for different spectral types are given in
Table~\ref{tab:10hr}, second column; 
\item Alternatively, one can assume that a proxy indicator for the total
luminosity of the network elements is the lowest Ca~II~HK fluxes seen in
main-sequence stars as a function of their spectral type. Noyes et al. (1984)
show in their Fig.~2 that ${\log \mathrm{R'(HK)}}$, i.e. the chromospheric
Ca~II~HK emission flux as a fraction of the bolometric luminosity, has a lower
envelope ranging from $-5.2$ for solar spectral type to $-4.9$ for M0V spectral
type. 

If all network elements have comparable brightness, the basal R'(HK) 
implies that there must be twice as many of them per unit bolometric 
luminosity on an M0 star as on the Sun. But the M0 star has $\approx1/16$ the 
solar bolometric luminosity, so the total number of elements on the M0 
star would be 8 times fewer than on the Sun. The Poissonian 
fluctuations in their contribution to the stellar flux would be $\sqrt{8}$ 
times less than on the Sun, but the stellar luminosity on which they are 
superimposed is 16 times less. Overall, the fluctuations ($\Delta f/f$) on the
M0  star could be $16/\sqrt{16/2} \approx 6$ times greater than $\Delta f/f$
for the Sun. The  estimates for other spectral types proceed in a similar way
and are reported in Table~\ref{tab:10hr}, third column. 
\end{enumerate}
\begin{table}[htbp]  
\centering
\begin{tabular}{ccc}
\hline
Sp & $\sigma_{\star,1}/\sigma_\odot$ & $\sigma_{\star,2}/\sigma_\odot$ \\ 
\hline
G2 &              1.0                &              1.0                \\ 
G5 &              1.3                &              1.2                \\ 
K0 &              2.5                &              2.0                \\ 
K5 &              6.3                &              3.5                \\ 
M0 &             16                  &              6.0                \\ 
\hline
\end{tabular}
\caption{Stellar 10 hour variability as estimated by method (1) and (2).}
\label{tab:10hr}
\end{table}
%
%
\section{Conclusion and discussion}
For the Sun, the 10 hour variability and 10 day one are decoupled indicating
different physical origins. For other stars, the same decoupling is expected.
The fact that a significant fraction of them are much more active than the Sun
on long periods does not imply a similar variability on 10 hour timescale. 
Estimating the latter is presently \emph{quite uncertain}, as well from the
theoretical point of view (very little is known about its physical origin) as
from the observational one (amplitudes are too small to be observed from the
ground). One can only conjecture about it. We have considered two such
``guess-estimates''. For spectral type and magnitude around which most telluric
planet detections are expected, G and K stars with $m_V = 15-16$, the 10 hour
variability would be 2--6 times that of the Sun, whereas the threshold
where the photometry in different colors is more efficient is 7--12 times it.
\emph{If these estimates are relevant}, the colored photometry would not provide
a significant advantage. 

On the other hand, if the colored information is obtained thanks to an
instrument that does not loose many photons, as a dichroic and two CCD sets or
a dispersive system (e.g. derived from the proposal by Chao \& Chi (\cite{chao}))
and a single CCD, photometry in white is still possible and as efficient as in
the absence of color information. Then, \emph{the Precaution Imperative stresses
the advantage of having the capability of photometry in different colors} in
case our present estimates of the 10 hour variability are incorrect, and stars
are significantly more variable than expected. The main possible disadvantage of
the corresponding instrument would be cost and complexity, at a level that has
to be estimated.

In conclusion, it is our opinion that it is worth studying practical
implementations of photometry in different colors for the Eddington mission
before taking irreversible decisions. 
%
%

%
%
\end{document}